%% file: main.tex
\documentclass[conference]{IEEEtran}
\IEEEoverridecommandlockouts

\usepackage{cite}
\usepackage{amsmath,amssymb,amsfonts}
\usepackage{amsmath,amssymb,amsthm,amsfonts,bm,bbm}
\usepackage{algorithmicx, algorithm, algpseudocode}
\usepackage{graphicx}
\usepackage{textcomp}
\usepackage{xcolor}
\usepackage{subcaption}
\usepackage{tikz}
\usetikzlibrary{automata, positioning, arrows,shapes.multipart,fit}

\usepackage{pgfplots}
\usepackage{pgfplotstable}
\pgfplotsset{ignore legend/.style={every axis legend/.code={\renewcommand\addlegendentry[2][]{}}}}

\def\BibTeX{{\rm B\kern-.05em{\sc i\kern-.025em b}\kern-.08em
    T\kern-.1667em\lower.7ex\hbox{E}\kern-.125emX}}
\begin{document}

\title{DRL-based Dolph-Tschebyscheff Beamforming in Downlink Transmission for Mobile Users
\thanks{Nancy Nayak and Kin K. Leung are funded by EPSRC grant EP/Y037197/1.}
}

\author{\IEEEauthorblockN{Nancy Nayak and Kin K. Leung}
\IEEEauthorblockA{\textit{Dept of Electrical and Electronic Engineering} \\
\textit{Imperial College London}\\
London, United Kingdom \\
\{n.nayak, kin.leung\}@imperial.ac.uk}
\and
\IEEEauthorblockN{Lajos Hanzo}
\IEEEauthorblockA{\textit{Electronics and Computer Science} \\
\textit{University of Southampton}\\
Southampton, United Kingdom \\
hanzo@soton.ac.uk}
}

\maketitle

\begin{abstract}
With the emergence of AI technologies in next-generation communication systems, machine learning plays a pivotal role due to its ability to address high-dimensional, non-stationary optimization problems within dynamic environments while maintaining computational efficiency. One such application is directional beamforming, achieved through learning-based blind beamforming techniques that utilize already existing radio frequency (RF) fingerprints of the user equipment obtained from the base stations and eliminate the need for additional hardware or channel and angle estimations. However, as the number of users and antenna dimensions increase, thereby expanding the problem's complexity, the learning process becomes increasingly challenging, and the performance of the learning-based method cannot match that of the optimal solution. In such a scenario, we propose a deep reinforcement learning-based blind beamforming technique using a learnable Dolph-Tschebyscheff antenna array that can change its beam pattern to accommodate mobile users. Our simulation results show that the proposed method can support data rates very close to the best possible values.
\end{abstract}


\section{Introduction}
The next generation communication system is supposed to support a large number of user equipments (UEs) with high data rates to enable massive communication and immersive user experiences. Technologies such as millimeter-wave (mmWave) communications and massive Multiple-Input Multiple-Output (MIMO) enable efficient spectrum usage and numerous connections,  making them ideal for dense urban areas and high data-demand applications. Massive MIMO systems with mmWave technologies harness MIMO's spatial resources and high data rates because of the large available bandwidth at mmWave frequency bands but face significant path and penetration losses, which can be mitigated by directional transmission and antenna beamforming.

Traditional beamforming methods include beam sweep \cite{akdeniz2014millimeter}; however, such a brute force search through the possible steering directions is both time-consuming and requires energy expenditure. Beam steering using location information, as highlighted in \cite{va2017inverse}, is a popular approach but incurs additional overhead due to the need for location data. A method leveraging the Extended Kalman Filter to track mmWave beams in mobile scenarios with moving UEs was proposed in \cite{liu2019ekf}. In \cite{zhang2019exploring}, a temporal-correlation-aided semi-exhaustive search algorithm for beam alignment was presented. Angle of Arrival (AoA) and Angle of Departure (AoD) estimation leverage the sparsity of mmWave channels to reduce the number of required measurements compared to beam sweeping. This problem is typically addressed through compressed sensing (CS) signal processing approaches \cite{myers2019message}, channel estimation methods utilizing low-rank techniques \cite{brighente2020estimation}, or machine learning techniques based on channel state information (CSI) \cite{yang2020deep}. In this work, we look at blind beamforming methods that do not use any CSI. Blind beamforming utilizes a stochastic gradient algorithm to update the beamforming vectors solely based on the received signal power \cite{jiang2010mimo}.

\begin{figure*}[!t]
    \centering
    \begin{subfigure}{.25\linewidth}
        \resizebox{\linewidth}{!}{
            \pgfplotstableread[col sep = comma]{./data/data_03UE_BS04x04_rateEvo.csv}\datatablethree
            \pgfplotstableread[col sep = comma]{./data/data_10UE_BS04x04_rateEvo.csv}\datatableten
        \tikzstyle{mark_style} = []

         \input{./figs/rateevo_template.tex}
        }
        \caption{$N_t = 4 \times 4$}
        \label{fig:rateevo_bs04x04}
    \end{subfigure}%
    \begin{subfigure}{.25\linewidth}
        \resizebox{\linewidth}{!}{
            \pgfplotstableread[col sep = comma]{./data/data_03UE_BS07x07_rateEvo.csv}\datatablethree
            \pgfplotstableread[col sep = comma]{./data/data_10UE_BS07x07_rateEvo.csv}\datatableten
            \tikzstyle{mark_style} = []
          \input{./figs/rateevo_template.tex}
        }
        \caption{$N_t = 7 \times 7$}
        \label{fig:rateevo_bs07x07}
    \end{subfigure}%
    \begin{subfigure}{.25\linewidth}
        \resizebox{\linewidth}{!}{
            \pgfplotstableread[col sep = comma]{./data/data_03UE_BS10x10_rateEvo.csv}\datatablethree
            \pgfplotstableread[col sep = comma]{./data/data_10UE_BS10x10_rateEvo.csv}\datatableten
        \input{./figs/rateevo_template.tex}
        }
        \caption{$N_t = 10 \times 10$}
        \label{fig:rateevo_bs10x10}
    \end{subfigure}%
    \caption{Rate Evolution during learning while using UPA.}
    \label{fig:rateevo}
\end{figure*}
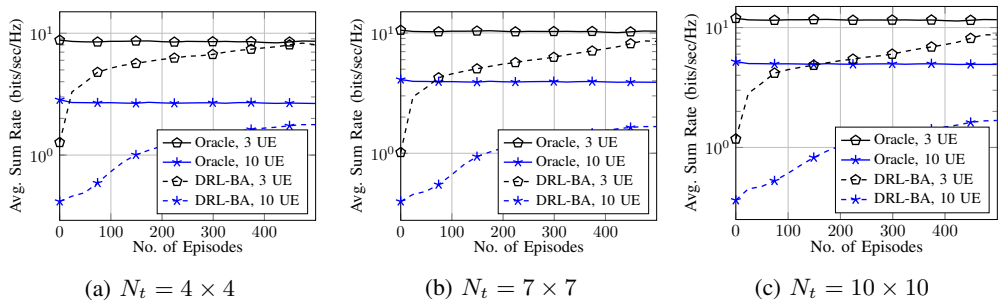

Deep Learning, including deep reinforcement learning (DRL), techniques are extensively used in mmWave communication systems. Examples include utilizing autoencoder-based models to improve hybrid precoding \cite{huang2019deep}, replacing hybrid precoding with deep learning-based approaches to predict optimal transmit/receive beam pairs from the observed channel characteristics \cite{li2019deep}, and leveraging DRL for beamforming tasks \cite{wang2020precodernet}. In \cite{raj2022deep}, the authors propose a DRL-based blind beam alignment (DRL-BA) method using only the RF fingerprint of UEs from base stations (BS) in a multi-BS cellular environment with multiple mobile UEs. This method learns to predict the best BS to serve each of the UEs and the corresponding beam alignment angles where the beam alignment relies on existing signals, eliminating the need for additional hardware like GPS or LIDAR. The authors compared their proposed method with an Oracle, which is assumed to have the exact knowledge of the best BS for every UE and the corresponding beam alignment angles. They show that the DRL-BA technique produces the same rate as an Oracle does in the case of small antenna arrays without any additional hardware/resources and without any labeled dataset for training. This method is suitable for even non-stationary environments and mobile UEs, unlike methods based on location, because location alone may not be sufficient for beamforming. In \cite{raj2022deep}, it is observed that larger antenna arrays result in a greater performance gap between the DRL-BA and the Oracle. This is because more antennas create sharper beams but also increase side lobes, leading to more inter-user interference. In this work, we aim to close this gap between the Oracle and DRL-BA by dynamically predicting the beamforming pattern of the antenna array based on the number of UEs each BS needs to serve.

Consider a BS equipped with a Uniform Planar Array (UPA) situated in the \textit{yz-plane} with $N_{t,h}$ and $N_{t,v}$ elements in $y$ and $z$ axis, respectively, and $N_t = N_{t,h} \times N_{t,v}$. In Fig. \ref{fig:rateevo}, we compare the evolution average sum rate (bits/sec/Hz) during learning for DRL-BA with the Oracle method for three different antenna array configurations of $N_t=4\times 4$, $N_t=7\times 7$ and $N_t=10\times 10$ in Fig. \ref{fig:rateevo_bs04x04}, \ref{fig:rateevo_bs07x07} and Fig. \ref{fig:rateevo_bs10x10}, respectively. One can observe that for a particular number of UEs, the data rate of Oracle increases with an increased number of antennas $N_t$ because as $N_t$ increases, the beam produced by the antenna array becomes narrower, and narrower beams can deliver most of the power towards the target direction. As Oracle has the knowledge of precise beam-alignment angles, with narrower beams, it causes the least amount of inter-user interference. As for the Oracle, the increase in data rate with an increase in antenna dimension is larger for $10$ UEs than for $3$ UEs because a narrower beam is helpful when the number of UEs is high. However, this is not very prominent for DRL-BA because, during learning with narrower beams, the target frequently falls outside the coverage area of the beam and, therefore, will incur high interference. Therefore, we observe that even though the data rates improve as $N_t$ increases for Oracle, corresponding improvement in data rates is not evident for DRL-BA. Further, this interference due to a narrower beam increases for additional UEs, creating a large performance gap in the case of $10$ UEs when compared to that with $3$ UEs, as shown in Fig. \ref{fig:rateevo_bs10x10}. Our goal here is to reduce this gap to support a large number of UEs and efficiently use additional antennas for mobile UEs.

In the majority of existing studies \cite{sohrabi2017hybrid, lota20175g, singh2015feasibility} including \cite{raj2022deep}, researchers predominantly focus on uniform antenna (UA) arrays, characterized by consistent spacing and amplitude among antenna elements. UAs are considered for delivering the lowest half-power beamwidth (HPBW), succeeding Dolph-Tschebyscheff (DT) and binomial arrays in performance. Although UA achieves the minimal HPBW and, consequently, the highest directivity, they are associated with relatively larger side lobes. The high directivity is essential for effective spatial multiplexing aimed at serving multiple UEs, implying a low HPBW; however, larger side lobes can exacerbate interference. Binomial arrays are known for their smaller side lobes but exhibit a broader HPBW, followed by DT and UA arrays. In scenarios where a BS is required to serve many UEs, minimizing side lobes is advantageous as it reduces interference. For a specified side-lobe level, the DT array achieves the smallest beamwidth between the first nulls \cite{balanis2015antenna}. Conversely, the DT design accomplishes the lowest possible side-lobe level for a given beamwidth between the first nulls. It becomes evident that designers are required to achieve a balance between side-lobe level and beamwidth based on the number of UEs that the BS is serving.

In this paper, we propose a DRL algorithm that learns to balance between the side-lobe level and the beamwidth in a DT antenna based on the number of mobile UEs served by each BS. The proposed method reduces the performance gap between DRL-BA and Oracle to a great extent. In the next section, we will first explain the system model and the idea of blind beam alignment before presenting the proposed method. Detailed simulation results will then be presented to reveal the merits of the new method.

\begin{figure*}[t]
\begin{equation}
\begin{aligned}
    \bm{a}_{t}(\phi,\theta, R) &=  \frac{1}{\sqrt{N_t}}\mathbf{s}_{t,h,m} \otimes \mathbf{s}_{t,v,n},
        \text{ where $\otimes$ is tensor product and } \\
        \mathbf{s}_{t,h,m} &=[a_{h,0}, \dots, a_{h,m}e^{j \frac{2\pi}{\lambda} dm \sin\phi \sin\theta }, \dots, a_{h,(N_{t,h}-1)e^{j \frac{2\pi}{\lambda} d (N_{t,h}-1) \sin\phi \sin\theta}}], \\
        \mathbf{s}_{t,v,n} &=[a_{v,0}, \dots, a_{v,n}e^{j \frac{2\pi}{\lambda} dn \cos \theta}, \dots, a_{v,(N_{t,v}-1)}e^{j \frac{2\pi}{\lambda} d (N_{t,v}-1) \cos \theta}],
    \label{eqn:resp_vec_proposed}
\end{aligned}
\end{equation}
\vspace{-5mm}
\end{figure*}
\begin{figure*}[t]
\begin{equation}
    \begin{aligned}
    \mathbf{a}_h &=[a_{h,0}, \dots, a_{h,(N_{t,h}/2-1)}, a_{h,(N_{t,h}/2-1)}, \dots,  a_{h,0}] \,\,
    \mathbf{a}_v =[a_{v,0}, \dots, a_{v,(N_{t,v}/2-1)}, a_{v,(N_{t,v}/2-1)}, \dots, a_{v,0}].
    \label{eqn:array}
\end{aligned}
\end{equation}
\vspace{-5mm}
\end{figure*}

\section{Blind mmWave Beam Alignment}
Consider a mmWave downlink communication system consisting of $N_{BS}$ BS serving $N_{UE}$ UEs using the same carrier frequency. Blind beamforming aims to select the best BS to serve each UE as well as the best set of beam alignment parameters for efficient beamforming. This is achieved by the central BS (or controller) selecting one BS that can best serve the UE based on the current information about UE (e.g., the received signal strength of beacon signal at UEs) and then selecting the right beam alignment directions. We assume a reliable link for data exchange between the BSs and the central BS and a dedicated control channel for additional signaling. 

Let each BS have a Planar Array (PA) antenna with $N_t$ transmit antenna elements. Each UE has $N_r = 1$ receive antenna and, therefore, omnidirectional transmission. Let the $i^{th}$ UE be served by the $j^{th}$ BS. If the signal transmitted for the $i^{th}$ UE from the $j^{th}$ BS is $x_i$, then the signal received at the $i^{th}$ UE is given by 
\begin{align}
    y_i &= P_{TX,j}\mathbf{H}_{ij} \mathbf{f}_{ij} x_i + \sum_{k=1, k\neq i}^{N_{UE}} P_{TX,j'} \mathbf{H}_{ij'} \mathbf{f}_{kj'} x_k + n,
\end{align}
where $j'=j$ if the $k^{th}$ and the $i^{th}$ UEs are served by the same BS. The channel $\bm{H}_{i,j}$ between the BS and each UE (of dimension $N_r \times N_t$) is modeled as Saleh-Valenzuela channel model \cite{el2014spatially}. The channel between the $j^{th}$ BS and the $i^{th}$ UE is characterized by $\bm{H}_{ij} = \sqrt{\frac{N_t N_r}{\kappa}}
            \sum \limits_{l=1}^{\kappa}
                \alpha_l
                    \bm{a}_r(\phi_l^r,\theta_l^r)
                    \bm{a}^*_t(\phi_l^t,\theta_l^t),$
where $\kappa$ is the number of propagation paths, $\alpha_l$ is the complex gain associated with $l^{th}$ path. $(\phi_l^t,\theta_l^t)$ are the azimuth and elevation angles of departure of $l^{th}$ ray at the BS, respectively. Similarly, $(\phi_l^r,\theta_l^r)$ are the angles of arrival of $l^{th}$ ray at the UE. We measure $\theta$ from \textit{+z-axis} and $\phi$ from \textit{+x-axis}. Sec. \ref{sec:proposedLDT} comprehensively discusses the UPA array coefficients and the unique aspects of the proposed method.

Since we assume omnidirectional reception with a single antenna at the receiver, $\bm{a}^{r}(\phi,\theta) = 1$. The path loss (in $dB$) of mmWave propagation is modeled as \cite{5gmodel2016}, $PL(f_c,d)_{dB} = 20 \log_{10} \left( \frac{4\pi f_c}{c} \right) 
        + 10 n \left(1 + b \left(\frac{f_c-f_0}{f_0}\right) \right)
        \log_{10} \left(d\right)
        + X_{\sigma dB},$
where $n$ is the path loss exponent, $f_c$ is the carrier frequency, $f_0$ is the fixed reference frequency, $b$ captures the frequency dependency of path loss exponent and $X_{\sigma dB}$ is the shadow fading term in $dB$. The Signal to Interference plus Noise Ratio (SINR) at $i^{th}$ UE, which is served by $j^{th}$ $\mu$BS, is given by
\begin{align}
    \zeta^{i} = \frac{P_{TX,j} |\bm{H}_{i,j} \bm{f}_{ij}|^2}
                {\sum \limits_{k=1, k \neq i}^{N_{UE}} P_{TX,j'} |\bm{H}_{ij'} \bm{f}_{kj'}|^2
                    + \sigma^2}, \label{eqn:ue_sinr}
\end{align}
where $\bm{H}_{i,j}$ is the channel between $i^{th}$ UE and $j^{th}$ BS, $\bm{f}_{ij}$ is the transmit codeword used by $j^{th}$ BS for $i^{th}$ UE, and $\sigma^2$ is the noise power and $P_{TX,j}$ is the transmit power of $j^{th}$ $\mu$BS. 

Each UE transmits a unique omnidirectional beacon signal, which is received in a faded form by all BSs. The BS obtains a noisy RF signature of all UEs via the control channel, eliminating the need for additional devices. Each UE transmits with a power of $p_u$ $dB$. The received signal strength indicator (RSSI) in $dB$ at the BS from a UE is calculated as 
\begin{equation}
    \label{eq:rssi}
    RSSI = p_u + g_{tx} - PL(f_c,d)_{dB} + g_{rx}
\end{equation}
where $g_{tx}$ is the gain of the transmit antenna at UE and $g_{rx}$ is the gain of the receive antenna at the BS, $f_c$ is the carrier frequency of the band over which RSSI is calculated \cite{5gmodel2016}.

\section{Proposed DRL-based DT antenna array}
\label{sec:proposedLDT}
The array response vector of transmitter, $\bm{a}_{t}(\phi,\theta)$ (of dimension $N_t = N_{t,h} \times N_{t,v}$) at the UPA of the BS, is given by $\bm{a}_{t}(\phi,\theta) = \frac{1}{\sqrt{N_t}} \left[ 
    1, 
    \ldots,
    e^{j \frac{2\pi}{\lambda} d \left(m \sin\phi \sin\theta + n \cos \theta \right)},
    \ldots, 
    \right. \nonumber \\ \left.
    e^{j \frac{2\pi}{\lambda} d \left((N_{t,h}-1) \sin\phi \sin\theta 
            + (N_{t,v}-1) \cos \theta \right)}
\right]^T,$ where $0 < m < N_{t,h} - 1$ and $0 < n < N_{t,v} - 1$ and $d$ is the inter-element spacing. Instead of using UPA, we propose to use \textit{Dolph-Tschebyscheff planar array} (DTPA) of $N_t = N_{t,h} \times N_{t,v}$ elements with spacing $d$ between the elements. We propose to train the DTPA because, unlike UPA, the antenna coefficients of DTPA can be tuned using a parameter called \textit{major-to-minor lobe ratio} based on the requirement. A DT antenna array at the $j^{th}$ BS is usually specified with a predefined major-to-minor lobe ratio or the voltage ratio $r_j (dB)$, indicating that the side lobe is $r_j$ dB below the maximum of the major lobe. Regarding the side lobe, a higher $r_j$ value corresponds to greater suppression of side lobes (i.e., lower side-lobe levels), improving the ability to focus energy in the main direction and reducing interference with others. However, in terms of sharpness, increasing $r_j$ results in a wider main lobe, reducing the sharpness or directivity of the main beam. Decreasing $r_j$ increases the sharpness of the beam (narrower the main lobe) but allows higher side-lobe levels. The DTPA response vector of transmitter, $\bm{a}_{t}(\phi,\theta, r_j)$ (of dimension $N_t$), can be given by (\ref{eqn:resp_vec_proposed}), where $\mathbf{a}_h$ and $\mathbf{a}_v$ can be found by the design procedure of the DT antenna array. Considering an even number of antenna elements in the planar array of the BS and assuming that the amplitude excitation is symmetrical about the origin, following \cite{balanis2015antenna}, the coefficients $\mathbf{a}_h$ and $\mathbf{a}_v$ of the array factors $\mathbf{s}_{t,h,m}$ and $\mathbf{s}_{t,h,n}$, respectively, can be rewritten as shown in (\ref{eqn:array}). These antenna coefficients are functions of $r_j$ for the $j^{th}$ BS, and the beam patterns can be adjusted by learning a proper $r_j$.

It is well-established that a reduction in the HPBW typically results in an increase in the side-lobe levels. In the presence of a large number of UEs per BS, the HPBW can be sacrificed for the sake of a lower co-channel interference level among the UEs served by the same BS. Therefore, in our system model, we use a DT antenna array instead of UPA so that the side lobes of the beam can be adjusted by learning an appropriate $r_j$ for every BS based on the number of UEs served by that BS. The complete procedure for learning the DTPA is outlined below. We first discuss the procedure for the DT linear array for $N_{t,h}$ antennas and then extend it to the DT planar array, i.e., for $N_t = N_{t,h} \times N_{t,v}$ antennas). To improve convergence during learning, we consider a DTPA where the array coefficients are given by $\mathbf{a}_h$ and $\mathbf{a}_v$ in (\ref{eqn:array}) are identical.

The DT linear array factor for $N_{t,h}$ number of antenna elements (where $N_{t,h}$ is even) is given by 
\begin{equation}
    (AF)_{N_{t,h}} = \sum\limits_{n=1}^{N_{t,h}/2} a_n \cos{(2n-1)u},
    \label{eq:arrayfactor}
\end{equation}
where $u = \frac{\pi d}{\lambda} \cos{\theta}$ and $a_n$ are the excitation coefficients, which are related to the Tschebyscheff polynomial. The array factor in (\ref{eq:arrayfactor}) is expanded and each $\cos(mu)$ function is replaced by the appropriate series expansion given by 
\begin{equation*}
    \begin{aligned}
        &m=0, \cos(mu)=1=T_0(z), \\
        &m=1,\cos(mu)=\cos(u)=z=T_1(z), \\
        &m=2,\cos(mu)=\cos(2u)=2\cos^2{u}-1=2z^2-1=T_2(z),
    \end{aligned}
\end{equation*}
where in general, $T_{m}(z) = 2zT_{m-1}(z) - T_{m-2}(z)$ \cite{balanis2015antenna}. We determine the point $z=z_0$ such that $T_m(z_0)=r_j$ for the $j^{th}$ BS. The design procedure requires that the Tschebyscheff polynomial in the $-1\leq z\leq z_1$, where $z_1$ is the null nearest to $Z=+1$, be used to represent the minor lobe of the array. The major lobe of the pattern is formed from the remaining part of the polynomial up to point $Z_0$. Substitute $\cos{u}=z/z_0$ in the array factor found in (\ref{eq:arrayfactor}) and we equate this to $T_{N_{t,h}-1}(z)$. From this, the array excitation coefficients $a_n$'s can be determined, and hence $a_n$'s are functions of the $r_j$. For more details, readers are referred to Example 6.9 on Page 335 of \cite{balanis2015antenna}. $\bm{r}$ is learned based on the spatial distribution of the UEs, and therefore, the proposed method is named the Learned DTPA (LDTPA). 

We model the problem of estimating $\mathbf{r}$ along with the identification of the best BS and the beam alignment angles as a Markov decision process (MDP). An MDP consists of a state space $\mathcal{S}$, an action space $\mathcal{A}$, an initial state distribution $p(s_1)$, a stationary distribution for state transitions, which obey Markov property $p(s_{t+1}|s_{t}, a_{t}) = p(s_{t+1}|s_{t}, a_{t}, \ldots, s_{1}, a_{1})$ and a reward function $r: \mathcal{S} \times \mathcal{A} \rightarrow \mathbb{R}$. The learning agent takes $s_t$ as a concatenation of feature vectors of $N_{UE}$ UEs as input. Each feature vector is of dimension $N_{BS} + 1$ and consists of a) the RF fingerprints ($N_{BS}$ elements) seen at the BS about the associated UE from the received beacon signal, and b) the SINR experienced by that UE with the current beam alignment configuration from the BS serving the UE. The action taken by the agent includes a discrete value referring to the index of the BS to serve each UE, the $(\phi,\theta)$ pair, and also $r_j$. As the action space is a mix of discrete BS selection and continuous beam alignment angles selection and voltage ratio, we use a neural architecture that can handle both pseudo-discrete and continuous action \cite{raj2022deep}. The reward at each time instant $t$, the average sum rate (bits/sec/Hz) is defined as $\gamma_t = \frac{1}{N_{UE}} \sum \limits_{i=1}^{N_{UE}} \log_2 \left( 1 + \zeta_t^i \right),$ where $\zeta_t^i$ is the instantaneous SINR at $i^{th}$ UE as defined in 
(\ref{eqn:ue_sinr}).

In this work, we use the Deep Deterministic Policy Gradient (DDPG) as the learning algorithm to train the DRL agent. DDPG has an actor network that predicts the action $a_t$ based on the current state $s_t$ and a critic network, which computes the Q-value for the predicted action $Q(s_t, a_t)$. The critic network evaluates the quality of actions and provides feedback that guides the actor towards taking the optimal actions. Simultaneously, the critic network refines its predictions by learning from the observed rewards after each action, calculating the Q-value, and minimizing the Mean Square Bellman Error (MSBE) in its predictions through gradient-based optimization \cite{raj2022deep}. DDPG maintains a replay buffer $\mathcal{B}$ to obtain stable and uncorrelated gradients for policy improvement. DDPG also uses target networks to avoid divergence in value estimation. 

\begin{figure}[t]
    \centering
    \resizebox{0.5\textwidth}{!}{ %
    \input{./figs/neural_arch.tex}
    }
    \caption{Architecture of Proposed Neural Action Predictor.}
    \vspace{-5mm}
    \label{fig:nn_arch}
\end{figure}
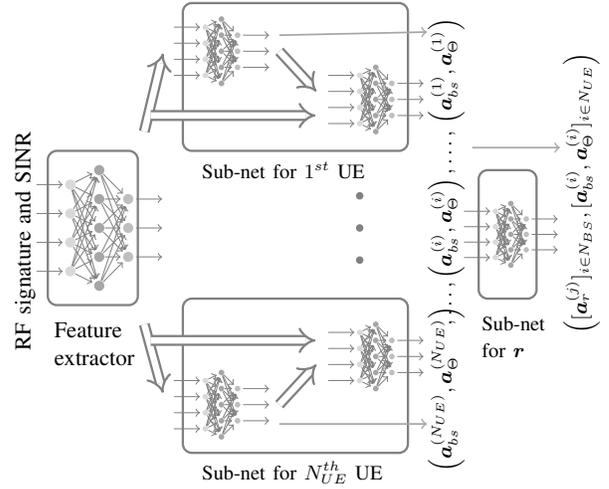

As the Q-value is continuous, a feed-forward neural network with a scalar output is used as the critic network. The proposed actor network has two parts - the first part of the architecture consists of $N_{UE}$ sub-networks (SNs), each one of which predicts the best BS for the UE and the corresponding beam alignment angles, is inspired by \cite{raj2022deep}; and the second part takes the outputs of the first part as input and predicts the voltage ratio. In the first part, all SNs share a common feature extractor (FE), which operates on the input to provide each UE SN with a set of features that can be used to select the action corresponding to that UE. The output of the first part, namely the best BS and beamforming angles, helps to predict and enhance the voltage ratio because depending on the number of UEs for a particular BS and the spatial placement of the UEs, the voltage ratio is set in such a way that the inter-user interference is minimized. The architecture of the proposed Neural Action Predictor is given in Fig. \ref{fig:nn_arch}.  

At $t^{th}$ time instant, the FE takes $\bm{x}_0 = s_t \in \mathbb{R}^{N_{UE}\cdot(N_{BS}+1)}$ as input and the output of $l^{th}$ layer of FE is $\bm{x}_{l} = g\left( \bm{W}_l \bm{x}_{l-1} + \bm{b}_l \right), \text{ for } l = 1, \ldots, L$ where $\bm{W}_l$ and $\bm{b}_l$ be the weight and bias associated with layer $l$, and $g(\cdot)$ is a non-linear activation function. The extracted features $\bm{x}_L$ is then fed to each of the SNs for action predictions for each UE.

Each SN includes a single layer for the BS prediction, with the actor SN for the $i^{th}$ UE predicting normalized scores across all BS indices via a softmax layer $\bm{a}^{(i)}_{bs} = \text{softmax}(\bm{W}_{i,bs} \bm{x}_{L} + \bm{b}_{i,bs}).$ Every element of the output $\bm{a}^{(i)}_{bs}$ is in the interval $[0,1]$ with the sum of components being $1$. The BS for serving the $i^{th}$ UE is selected as the one with the highest normalized score in $\bm{a}^{(i)}_{bs}$. Since the proposed method learns the voltage ratio $\bm{r}$ as well, exploring the optimal BS for each UE by adding noise is not necessary, which, in a way, decelerates convergence.

The elevation and azimuth angles for beam alignment depend on the UE position information available via $\bm{x}_L$ and the selected BS $\bm{a}^{(i)}_{bs}$. To integrate these, a fusion layer is used, and the input to this layer is a concatenated feature vector $\bm{y}_{i} = [\bm{x}_L, \bm{a}^{(i)}_{bs}]$ for each UE. The beam alignment actions for the $i^{th}$ UE are then computed as $\bm{a}^{(i)}_{\Theta} = \text{tanh}(\bm{W}_{i,\Theta} \bm{y}_i,  + \bm{b}_{i,\Theta}),$ where tanh($\cdot$) activation function outputs values in range $[-1, +1]$ and hence $\bm{a}^{(i)}_{\Theta} \in [-1, +1]^2$. To explore the space of elevation and azimuth angles, random Gaussian noise $n=\mathcal{N}(0,\sigma_{\Theta})$ is added to $\bm{a}^{(i)}_{\Theta}$ where the variance $\sigma_{\Theta}$ decays linearly with time. The modified output from the actor SN regarding the elevation and azimuth angles is given by $\Tilde{\bm{a}}^{(i)}_{\Theta} = \bm{a}^{(i)}_{\Theta}+n.$

Noting that the voltage ratio of every BS depends on the number of UEs served by the BS and the beam alignment angles, the two vectors $[\bm{a}^{(i)}_{bs}]_{i\in N_{UE}}$ and $[\Tilde{\bm{a}}^{(i)}_{\Theta}]_{i\in N_{UE}}$ are concatenated to form $\bm{z}= \left[[\bm{a}^{(i)}_{bs}]_{i\in N_{UE}},[\Tilde{\bm{a}}^{(i)}_{\Theta}]_{i\in N_{UE}}\right]$ and this fused information is passed via a layer. Then the action corresponding to the voltage ratio is given by $\bm{a}^{(j)}_{r} = \text{tanh}(\bm{W}_{r} \bm{z},  + \bm{b}_{r}),$ where $\bm{W}_{r}$ and $\bm{b}_{r}$ are weights and biases of the SN for predicting $\bm{r}$. Due to the use of tanh($\cdot$) activation, $\bm{a}^{(i)}_{r}\in [-1,+1]^{N_{BS}}$. To explore the space of $\bm{r}$, $\mathcal{N}(0,\sigma_{r})$ is added to $\bm{a}^{(j)}_{r}$ where $\sigma_{r}$ decays linearly with time. The modified output from the actor SN regarding $\bm{r}$ is given by $\Tilde{\bm{a}}^{(j)}_{r} = \bm{a}^{(j)}_{r}+n.$ The elevation and azimuth angles for beam alignment are computed (in radians) as
$
\theta_i = \frac{3\pi}{4} + \Tilde{\bm{a}}^{(i)}_{\Theta}[1] \times \frac{\pi}{4},
            \label{eqn:compute_theta} \text{ and }
\phi_i = \Tilde{\bm{a}}^{(i)}_{\Theta}[2] \times \frac{\pi}{2}.
            \label{eqn:compute_phi}
$
The computation for elevation angle is based on the assumption that all UEs are below the height of $\mu$BS and hence $\theta_i \in [\pi/2, \pi]$. Similarly, it is assumed that $\phi_i \in [-\pi/2, +\pi/2]$. The voltage ratio is computed by $r_j = R_0\times(\Tilde{\bm{a}}^{(j)}_{r}+p)/q$, where $p=1.5,\, q=2.5$ and $R_0=40$, so that resultant $r_j \in [8,40]$. If we have prior knowledge about this range, $p$, $q$, and $R_0$ can be chosen accordingly. The proposed LDTPA algorithm is presented in Alg. \ref{alg:proposed_beamer}.

\begin{algorithm}[!t] 
\caption{Proposed algorithm for LDTPA}
\begin{algorithmic}[1]
    \State Initialize the actor, the critic, and the target networks 
    \State Create an empty replay buffer $\mathcal{B} \leftarrow \{\}$ with size $\tau$.
    \For {episode $= 1 \ldots M$}
        \State Select a random valid action for each UE.
        \State Observe the SINR and RF signature of each UE as $s_t$.
        \For {$t = 1 \ldots T$}
            \State Set $\bm{a}^{(i)}_{bs}$, $\Tilde{\bm{a}}^{(i)}_{\Theta}$ and $\Tilde{\bm{a}}^{(j)}_{r}$.
            \State Get new state observation $s_{t+1}$.
            \State Extract individual SINR for each UE from $s_{t+1}$ and compute average rate as reward $r_t$.            
            \State $\mathcal{B} \gets \mathcal{B} \cup (s_t, a_t, r_t, s_{t+1})$.
            \If {$|\mathcal{B}| \geq \tau$}
                \State Delete oldest experience from $\mathcal{B}$.
            \EndIf
            \State Sample $N$ experiences $(s_i, a_i, r_i, s_{i+1})$ from $\mathcal{B}$.
            \State Compute return followed by computing MSBE. 
            \State Update actor, critic, and the target networks.
        \EndFor
    \EndFor
\end{algorithmic}
\label{alg:proposed_beamer}
\end{algorithm}

\section{Simulation results and discussion}
We consider a four-junction scenario \cite{alkhateeb2018deep}, with $N_{BS} = 10$ and an inter-cell radius of approximately $100m$. We assume $f_c = 28 GHz$ with a bandwidth of $5 MHz$. Each BS is assumed to have a UPA antenna with $d = \lambda/2$, where $\lambda$ is the wavelength associated with $f_c$. Following the Street Canyon configuration \cite{5gmodel2016}, we use $n = 1.98$, $\sigma=3.1$, $b = 0$ and $f_0 = 10^{9}$. The RF signature is calculated by eq. (\ref{eq:rssi}) uses $f_c=28GHz$ for its path loss calculation. We set $N_{UE} = 10$, and $\kappa = 1$. We simulate channel evolution using a Markov Chain Monte Carlo method, updating UE locations and drawing channel samples at each timestep. The performance of the proposed method is compared with the following baseline methods:
\begin{itemize}
    \item Oracle: Equipped with the exact knowledge about the location of UEs, this agent picks the best serving BS for each UE as well as the corresponding beam alignment angles, and the inter-user interference is minimized. This is the maximum expected data rate any algorithm can achieve.
    \item DRL-BA: DRL-BA has a feed-forward critic network, and UE SN augmented actor-network with $L = 2$. Each hidden layer has $128$ nodes. We provide the results averaged over $4$ agents, each trained for $500$ episodes. Each episode consists of $1,000$ time steps, and the UE positions are reset at the end of each episode. For mobile UEs, some BSs may need to serve more UEs than others, which incurs more inter-user interference. So, the beam patterns need to be adjusted dynamically based on the number of UEs that a BS serves. DRL-BA uses the UPA for higher directivity, but for increased antenna configuration, the number of side lobes increases, resulting in higher inter-user interference. 
    \item DTPA: Compared to UPA, the DT antenna array has adjustable main-lobe width where the side-lobe suppression is balanced with directivity \cite{balanis2015antenna}. In the experiments, the DT antenna array is chosen with $r_j=26dB$. 
    \item Proposed LDTPA: We propose to learn the voltage ratio $\bm{r}$ and predict its close optimal value online using the DRL algorithm (apart from predicting the best BS-UE assignment and the beam alignment angles), depending on the required directivity and HPBW to serve a certain number of UEs. For a BS serving many UEs, the proposed method should learn to predict a higher value of $r_j$ for the $j^{th}$ BS to have lower side-lobes beams, and the UEs have less inter-user interference. However, for a BS serving a small number of UEs, it can have less interference even with additional side lobes and still achieve high data rates. Furthermore, learning beam alignment angles of a wider beam width improve the learning convergence.  
\end{itemize}

Fig. \ref{fig:rateevo_10ue_bs08x08} and Fig. \ref{fig:rateevo_10ue_bs10x10} show the time evolution of the average sum rate of the UEs for antenna configurations $N_t=8\times 8$ and $N_t=10\times 10$, respectively. In both cases, LDTPA reduces the performance gap and almost reaches the highest data rate of the Oracle method. The DTPA supports a larger data rate than UPA-based DRL-BA because of a reduced side-lobe level for $r_j=26$ dB in DTPA for all the BSs. However, in the presence of mobile users, the LDTPA learns to predict an appropriate $r_j$ for $j^{th}$ BS based on user positions and their channel qualities inherently. At the beginning of the training, as the $r_j$'s have not been learned yet, LDTPA has a similar data rate as DTPA. However, over time, as the $r_j$'s are learned, the data rate increases, and the gap between the DRL-BA and Oracle reduces, indicating that the agent learns to predict the pattern of the beam well by the proposed LDTPA method. The cumulative distribution of the data rate achieved is given in Fig. \ref{fig:ratecdf_10ue_bs10x10}. We simulate $10,000$ observations ($10$ episodes), and the data aggregated based on these observations are plotted. Note that the trained DRL agents are used to get the rate distribution. The proposed LDTPA reaches very close to the optimal values by the Oracle, even in the case of a larger antenna array configuration of $N_t=10\times10$.{With a batch size of $N$, the complexity of training the proposed actor-network of LDTPA has a complexity of $\mathcal{O}(N . N_{UE}(N_{BS}+1) . d + N.d^2)$ for FE, $\mathcal{O}(N. N_{UE}.(d.N_{BS}+(d+N_{BS}).2)$ for $\bm{a}^{(i)}_{bs}$ and $\Tilde{\bm{a}}^{(i)}_{\Theta}$ together; and $\mathcal{O}(N. N_{UE}(N_{BS}+2). N_{BS})$ for $\Tilde{\bm{a}}^{(j)}_{r}$.

\begin{figure}[!t]
\centering
\begin{subfigure}{.48\linewidth}
    \resizebox{\linewidth}{!}{
        \pgfplotstableread[col sep = comma]{./data/data_10UE_BS08x08_rateEvo_compare.csv}\datatableten
    \input{./figs/rateevo_template_compare.tex}
    }
    \caption{$N_t = 8 \times 8$}
    \label{fig:rateevo_10ue_bs08x08}
\end{subfigure}%
\begin{subfigure}{.48\linewidth}
    \resizebox{\linewidth}{!}{
        \pgfplotstableread[col sep = comma]{./data/data_10UE_BS10x10_rateEvo_compare.csv}\datatableten
    \input{./figs/rateevo_template_compare.tex}
    }
    \caption{$N_t = 10 \times 10$}
    \label{fig:rateevo_10ue_bs10x10}
\end{subfigure}
\caption{Rate Evolution during learning}
\end{figure}
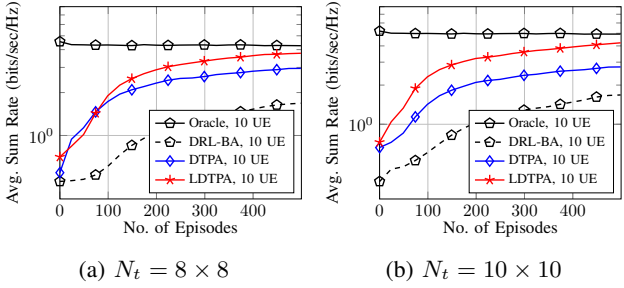

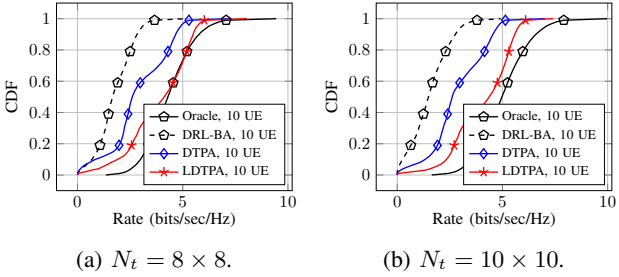
\begin{figure}[!t]
    \centering
    \begin{subfigure}{.48\linewidth}
        \resizebox{\linewidth}{!}{
            \pgfplotstableread[col sep = comma]{./data/data_10UE_BS08x08_rateCdf_compare.csv}\datatable
            \input{./figs/ratecdf_template_compare.tex}
        }
    \caption{$N_t = 8 \times 8$.}
    \label{fig:ratecdf_10ue_bs08x08}
    \end{subfigure}%
    \begin{subfigure}{.48\linewidth}
        \resizebox{\linewidth}{!}{
            \pgfplotstableread[col sep = comma]{./data/data_10UE_BS10x10_rateCdf_compare.csv}\datatable
            \input{./figs/ratecdf_template_compare.tex}
        }
    \caption{$N_t = 10 \times 10$.}
    \label{fig:ratecdf_10ue_bs10x10}
    \end{subfigure}
    \caption{CDF of observed rates}
    \vspace{-5mm}
\end{figure}

\section{Conclusion}
This paper investigates a blind beam alignment problem for mobile UEs in a multi-BS environment. First, we show that a DRL agent employing a UPA at the BS cannot produce a data rate as high as the Oracle. Then, we demonstrate that LDTPA closes the performance gap by optimizing beam patterns based on the number of UEs per BS as it learns to predict the voltage ratio dynamically, alongside identifying the best BS and beamforming angles. This efficient beamforming approach enhances ISAC sensing and boosts security through angular selectivity rather than omni-transmission. Our simulation results have shown that the proposed method can support data rates very close to the best possible values.
\bibliographystyle{IEEEtran}
\bibliography{library.bib}
\end{document}

%% file: figs/rateevo_template.tex
\begin{tikzpicture}[thick,scale=0.8]
    \begin{semilogyaxis}[
        width=7cm,
        height=6cm,
        xmin=0,
        xmax=500,
        ymin=0.0,
        ymax=15,
        grid=major,
        xlabel={No. of Episodes},
        ylabel={Avg. Sum Rate (bits/sec/Hz)},
        xlabel style={at={(0.50,0.05)}},
        ylabel style={at={(0.06,0.50)}},
        ytick={0,1,10},
        xtick={0,100, 200, 300, 400},
        legend pos=south east,
        legend cell align={left},
        legend style={fill opacity=0.6, draw opacity=1.0, text opacity=1.0, font=\small}
        ]
        

        \addplot[black, solid, thick, mark=pentagon, mark size={3.0}, mark repeat=3,
                ] 
            table [x=x_data, y expr=\thisrow{y_greedy}/1000, col sep=comma]{\datatablethree};
        \addlegendentry{Oracle, 3 UE};

        \addplot[blue, solid, thick, mark=star, mark size={3.0}, mark repeat=3,
                ] 
            table [x=x_data, y expr=\thisrow{y_greedy}/1000, col sep=comma]{\datatableten};
        \addlegendentry{Oracle, 10 UE};
        


        \addplot[black, dashed, thick, 
                mark=pentagon, mark options={solid}, mark size={3.0}, mark size={3.0}, mark repeat=3,
                ] 
            table [x=x_data, y expr=\thisrow{y_ddpg_2x128_gamma060}/1000, col sep=comma]{\datatablethree};
        \addlegendentry{DRL-BA, 3 UE};

        \addplot[blue, dashed, thick, 
                mark=star, mark options={solid}, mark size={3.0}, mark size={3.0}, mark repeat=3,
                ] 
            table [x=x_data, y expr=\thisrow{y_ddpg_2x128_gamma060}/1000, col sep=comma]{\datatableten};
        \addlegendentry{DRL-BA, 10 UE};
        
        
       
    \end{semilogyaxis}
\end{tikzpicture}

%% file: figs/neural_arch.tex
\tikzset{
    pics/ann/.style={
        code={
            \tikzstyle{every pin edge}=[<-,shorten <=1pt]
            \tikzstyle{neuron}=[circle,fill=black!25,minimum size=#1*10pt,inner sep=0pt]
            \tikzstyle{input neuron}=[neuron, fill=gray!30, pin=left:];
            \tikzstyle{output neuron}=[neuron, fill=gray!50, pin={[pin edge={->}]right:}];
            \tikzstyle{hidden neuron}=[neuron, fill=gray!70];
            
            \node[input neuron] (I1) at (-1.0*#1, -1.5*#1) {};
            \node[input neuron] (I2) at (-1.0*#1, -0.5*#1) {};
            \node[input neuron] (I3) at (-1.0*#1, +0.5*#1) {};
            \node[input neuron] (I4) at (-1.0*#1, +1.5*#1) {};

            \node[hidden neuron] (H1) at (0.0*#1, -2.0*#1) {};
            \node[hidden neuron] (H2) at (0.0*#1, -1.0*#1) {};
            \node[hidden neuron] (H3) at (0.0*#1, +0.0*#1) {};
            \node[hidden neuron] (H4) at (0.0*#1, +1.0*#1) {};
            \node[hidden neuron] (H5) at (0.0*#1, +2.0*#1) {};

            \node[output neuron] (O1) at (1.0*#1, -1.0*#1) {};
            \node[output neuron] (O2) at (1.0*#1, +0.0*#1) {};
            \node[output neuron] (O3) at (1.0*#1, +1.0*#1) {};

            \foreach \source in {1,...,4}
                \foreach \dest in {1,...,5}
                    \path (I\source) edge (H\dest);
            
            \foreach \source in {1,...,5}
                \foreach \dest in {1,...,3}
                    \path (H\source) edge (O\dest);
        }
    }
}

\tikzstyle{block} = [rectangle, draw, 
    text width=8em, text centered, rounded corners, minimum height=4em]

\begin{tikzpicture}[shorten >=1pt,->,draw=black!50]

    
    \draw[rounded corners,line width=1pt] (1.4,2.3) rectangle (2.8,4.7);
    \pic at (2.2,3.5) {ann=0.45};
    \node[rotate=90] (input) at (1.05,3.3) {RF signature and SINR};
    \node[text width=2cm] (initial) at (2.5,1.70) {Feature \\ extractor};
    
    \draw[rounded corners,line width=1pt] (3.5,0) rectangle (7.0,2.4);
    \pic at (4.10,0.75) {ann=0.25};
    \pic at (6.50,1.50) {ann=0.25};
    \node[rotate=90] (ue1_out) at (7.6,1.00) {\small $\left(\bm{a}_{bs}^{(N_{UE})}, \bm{a}_{\Theta}^{(N_{UE})}\right)$};
    \node[text width=4cm] (initial) at (5.80,-0.30) {\small Sub-net for $N_{UE}^{th}$ UE};
    
    \draw[rounded corners,line width=1pt] (3.5,4.7) rectangle (7.0,7.0);
    \pic at (4.10,6.25) {ann=0.25};
    \pic at (6.50,5.50) {ann=0.25};
    \node[rotate=90] (ueN_out) at (7.6,5.95) {\small $\left(\bm{a}_{bs}^{(1)}, \bm{a}_{\Theta}^{(1)}\right)$};
    \node[text width=4cm] (initial) at (5.80,4.40) {\small Sub-net for $1^{st}$ UE};

    \node[rotate=90] (ueN_out) at (7.6,3.5) {\small $,\dots,\left(\bm{a}_{bs}^{(i)}, \bm{a}_{\Theta}^{(i)}\right), \dots,$};
    
    \draw[line width=1pt,-implies,double, double distance=1mm] (2.9,2.0) -- (3.2,0.9);
    \draw[line width=1pt,-implies,double, double distance=1mm] (3.0,1.75) -- (5.6,1.75);
    \draw[line width=1pt,-implies,double, double distance=1mm] (5.00,0.75) -- (5.6,1.40);
    \draw[line width=0.75pt, gray!70] (5.00,0.45) -- (7.33,0.45);

    \draw[line width=1pt,-implies,double, double distance=1mm] (2.9,5.0) -- (3.2,6.25);
    \draw[line width=1pt,-implies,double, double distance=1mm] (3.0,5.25) -- (5.6,5.25);
    \draw[line width=1pt,-implies,double, double distance=1mm] (5.00,6.25) -- (5.6,5.60);
    \draw[line width=0.75pt, gray!70] (5.0,6.5) -- (7.33,6.55);

    \node[draw,shape=circle,fill=gray,minimum size=1mm,inner sep=0pt] (d1) at (6.25,4.0) {};
    \node[draw,shape=circle,fill=gray,minimum size=1mm,inner sep=0pt] (d2) at (6.25,3.5) {};
    \node[draw,shape=circle,fill=gray,minimum size=1mm,inner sep=0pt] (d3) at (6.25,3.0) {};


    \draw[rounded corners,line width=1pt] (8.1,2.4) rectangle (9.0,4.4);
    \pic at (8.6,3.4) {ann=0.240};
    \node[rotate=90] (ue1_out) at (9.7,4.05) {\small $\left([\bm{a}_{r}^{(j)}]_{i\in N_{BS}}, [\bm{a}_{bs}^{(i)}, \bm{a}_{\Theta}^{(i)}]_{i\in N_{UE}}\right)$};
    \node[text width=2.5cm] (initial) at (9.4,1.8) {\small Sub-net \\ for $\bm{r}$};
    \draw[line width=0.75pt, gray!70] (8.0,4.75) -- (9.4,4.75);

\end{tikzpicture}

%% file: figs/rateevo_template_compare.tex
\begin{tikzpicture}[thick,scale=0.98]
    \begin{semilogyaxis}[
        width=6.5cm,
        height=5.5cm,
        xmin=0,
        xmax=500,
        ymin=0.0,
        ymax=8,
        grid=major,
        xlabel={No. of Episodes},
        ylabel={Avg. Sum Rate (bits/sec/Hz)},
        xlabel style={at={(0.50,0.05)}},
        ylabel style={at={(0.06,0.50)}},
        ytick={0,1,10},
        xtick={0,100, 200, 300, 400},
        legend pos=south east,
        legend cell align={left},
        legend style={fill opacity=0.6, draw opacity=1.0, text opacity=1.0, font=\footnotesize}
        ]
        


        \addplot[black, solid, thick, mark=pentagon, mark size={3.0}, mark repeat=3,
                ] 
            table [x=x_data, y expr=\thisrow{y_greedy}/1000, col sep=comma]{\datatableten};
        \addlegendentry{Oracle, 10 UE};
        



        \addplot[black, dashed, thick, 
                mark=pentagon, mark options={solid}, mark size={3.0}, mark size={3.0}, mark repeat=3,
                ] 
            table [x=x_data, y expr=\thisrow{y_ddpg_2x128_gamma060}/1000, col sep=comma]{\datatableten};
        \addlegendentry{DRL-BA, 10 UE};

        \addplot[blue, solid, thick, 
                mark=diamond, mark size={3.0}, mark size={3.0}, mark repeat=3,
                ] 
            table [x=x_data, y expr=\thisrow{y_ddpg_2x128_gamma060_DT}/1000, col sep=comma]{\datatableten};
        \addlegendentry{DTPA, 10 UE};

        \addplot[red, solid, thick, 
                mark=star, mark size={3.0}, mark size={3.0}, mark repeat=3,
                ] 
            table [x=x_data, y expr=\thisrow{y_ddpg_2x128_gamma060_trainedDT_R3}/1000, col sep=comma]{\datatableten};
        \addlegendentry{LDTPA, 10 UE};

        
       
    \end{semilogyaxis}
\end{tikzpicture}

%% file: figs/ratecdf_template_compare.tex
\begin{tikzpicture}[thick,scale=0.98]
    \begin{axis}[
        width=6.5cm,
        height=5.5cm,
        xmin=-1.0,
        xmax=10.2,
        ymin=-0.1,
        ymax=+1.1,
        grid=major,
        xlabel={Rate (bits/sec/Hz)},
        ylabel={CDF},
        xlabel style={at={(0.50,0.05)}},
        ylabel style={at={(0.06,0.50)}},
        ytick={0.0,0.2,...,1.0},
        legend pos=south east,
        legend cell align={left},
        legend style={fill opacity=0.6, draw opacity=1.0, text opacity=1.0, font=\footnotesize}
        ]

        \addplot[black, solid, thick, mark=pentagon, mark size={3.0},
        mark size={3.0}, mark repeat=20, mark phase=20
                ] 
            table [y=cdf, x=greedy, col sep=comma]{\datatable};
        \addlegendentry{Oracle, 10 UE};

        \addplot[black, dashed, thick, 
                mark=pentagon, mark size={3.0}, mark size={3.0}, mark options={solid}, mark repeat=20, mark phase=20
                ] 
            table [y=cdf, x=ddpg_2x128_gamma060_000, col sep=comma]{\datatable};
        \addlegendentry{DRL-BA, 10 UE};

        \addplot[blue, solid, thick, 
                mark=diamond, mark size={3.0}, mark size={3.0}, mark repeat=20, mark phase=20
                ] 
            table [y=cdf, x=ddpg_2x128_gamma060_DT_000, col sep=comma]{\datatable};
        \addlegendentry{DTPA, 10 UE};
        
        \addplot[red, solid, thick, 
                mark=star, mark size={3.0}, mark size={3.0}, mark repeat=20, mark phase=20
                ] 
            table [y=cdf, x=ddpg_2x128_gamma060_trainedDT_R3_000, col sep=comma]{\datatable};
        \addlegendentry{LDTPA, 10 UE};

    \end{axis}
\end{tikzpicture}

%% file: main.bbl
\begin{thebibliography}{10}
\providecommand{\url}[1]{#1}
\csname url@samestyle\endcsname
\providecommand{\newblock}{\relax}
\providecommand{\bibinfo}[2]{#2}
\providecommand{\BIBentrySTDinterwordspacing}{\spaceskip=0pt\relax}
\providecommand{\BIBentryALTinterwordstretchfactor}{4}
\providecommand{\BIBentryALTinterwordspacing}{\spaceskip=\fontdimen2\font plus
\BIBentryALTinterwordstretchfactor\fontdimen3\font minus
  \fontdimen4\font\relax}
\providecommand{\BIBforeignlanguage}[2]{{%
\expandafter\ifx\csname l@#1\endcsname\relax
\typeout{** WARNING: IEEEtran.bst: No hyphenation pattern has been}%
\typeout{** loaded for the language `#1'. Using the pattern for}%
\typeout{** the default language instead.}%
\else
\language=\csname l@#1\endcsname
\fi
#2}}
\providecommand{\BIBdecl}{\relax}
\BIBdecl

\bibitem{akdeniz2014millimeter}
M.~R. Akdeniz, Y.~Liu, M.~K. Samimi, S.~Sun, S.~Rangan, T.~S. Rappaport, and
  E.~Erkip, ``Millimeter wave channel modeling and cellular capacity
  evaluation,'' \emph{IEEE journal on selected areas in communications},
  vol.~32, no.~6, pp. 1164--1179, 2014.

\bibitem{va2017inverse}
V.~Va, J.~Choi, T.~Shimizu, G.~Bansal, and R.~W. Heath, ``Inverse multipath
  fingerprinting for millimeter wave v2i beam alignment,'' \emph{IEEE
  Transactions on Vehicular Technology}, vol.~67, no.~5, pp. 4042--4058, 2017.

\bibitem{liu2019ekf}
F.~Liu, P.~Zhao, and Z.~Wang, ``{EKF}-based beam tracking for mmwave {MIMO}
  systems,'' \emph{{IEEE} Communications Letters}, vol.~23, no.~12, pp.
  2390--2393, 2019.

\bibitem{zhang2019exploring}
D.~Zhang, A.~Li, M.~Shirvanimoghaddam, Y.~Li, and B.~Vucetic, ``Exploring
  aoa/aod dynamics in beam alignment of mobile millimeter wave mimo systems,''
  \emph{IEEE Transactions on Vehicular Technology}, vol.~68, no.~6, pp.
  6172--6176, 2019.

\bibitem{myers2019message}
N.~J. Myers and R.~W. Heath, ``Message passing-based joint cfo and channel
  estimation in mmwave systems with one-bit adcs,'' \emph{IEEE Transactions on
  Wireless Communications}, vol.~18, no.~6, pp. 3064--3077, 2019.

\bibitem{brighente2020estimation}
A.~Brighente, M.~Cerutti, M.~Nicoli, S.~Tomasin, and U.~Spagnolini,
  ``Estimation of wideband dynamic mmwave and thz channels for 5g systems and
  beyond,'' \emph{IEEE Journal on Selected Areas in Communications}, vol.~38,
  no.~9, pp. 2026--2040, 2020.

\bibitem{yang2020deep}
Y.~Yang, F.~Gao, Z.~Zhong, B.~Ai, and A.~Alkhateeb, ``Deep transfer
  learning-based downlink channel prediction for fdd massive mimo systems,''
  \emph{IEEE Transactions on Communications}, vol.~68, no.~12, pp. 7485--7497,
  2020.

\bibitem{jiang2010mimo}
M.~Jiang, G.~Yue, and S.~Rangarajan, ``Mimo transmission with rank adaptation
  for multi-gigabit 60ghz wireless,'' in \emph{2010 IEEE Global
  Telecommunications Conference GLOBECOM 2010}.\hskip 1em plus 0.5em minus
  0.4em\relax IEEE, 2010, pp. 1--5.

\bibitem{huang2019deep}
H.~Huang, Y.~Song, J.~Yang, G.~Gui, and F.~Adachi, ``Deep-learning-based
  millimeter-wave massive {MIMO} for hybrid precoding,'' \emph{{IEEE}
  Transactions on Vehicular Technology}, vol.~68, no.~3, pp. 3027--3032, 2019.

\bibitem{li2019deep}
X.~Li and A.~Alkhateeb, ``Deep learning for direct hybrid precoding in
  millimeter wave massive mimo systems,'' in \emph{2019 53rd Asilomar
  Conference on Signals, Systems, and Computers}.\hskip 1em plus 0.5em minus
  0.4em\relax IEEE, 2019, pp. 800--805.

\bibitem{wang2020precodernet}
Q.~Wang, K.~Feng, X.~Li, and S.~Jin, ``Precodernet: Hybrid beamforming for
  millimeter wave systems with deep reinforcement learning,'' \emph{IEEE
  Wireless Communications Letters}, vol.~9, no.~10, pp. 1677--1681, 2020.

\bibitem{raj2022deep}
V.~Raj, N.~Nayak, and S.~Kalyani, ``Deep reinforcement learning based blind
  mmwave mimo beam alignment,'' \emph{IEEE Transactions on Wireless
  Communications}, vol.~21, no.~10, pp. 8772--8785, 2022.

\bibitem{sohrabi2017hybrid}
F.~Sohrabi and W.~Yu, ``Hybrid analog and digital beamforming for mmwave ofdm
  large-scale antenna arrays,'' \emph{IEEE Journal on Selected Areas in
  Communications}, vol.~35, no.~7, pp. 1432--1443, 2017.

\bibitem{lota20175g}
J.~Lota, S.~Sun, T.~S. Rappaport, and A.~Demosthenous, ``5g uniform linear
  arrays with beamforming and spatial multiplexing at 28, 37, 64, and 71 ghz
  for outdoor urban communication: A two-level approach,'' \emph{IEEE
  Transactions on vehicular technology}, vol.~66, no.~11, pp. 9972--9985, 2017.

\bibitem{singh2015feasibility}
J.~Singh and S.~Ramakrishna, ``On the feasibility of codebook-based beamforming
  in millimeter wave systems with multiple antenna arrays,'' \emph{IEEE
  transactions on Wireless Communications}, vol.~14, no.~5, pp. 2670--2683,
  2015.

\bibitem{balanis2015antenna}
C.~A. Balanis, \emph{Antenna theory: analysis and design}.\hskip 1em plus 0.5em
  minus 0.4em\relax John wiley \& sons, 2015.

\bibitem{el2014spatially}
O.~El~Ayach, S.~Rajagopal, S.~Abu-Surra, Z.~Pi, and R.~W. Heath, ``Spatially
  sparse precoding in millimeter wave {MIMO} systems,'' \emph{{IEEE}
  transactions on wireless communications}, vol.~13, no.~3, pp. 1499--1513,
  2014.

\bibitem{5gmodel2016}
``{5G Channel Model for bands up to 100 GHz},'' Tech. Rep. October, 2016.

\bibitem{alkhateeb2018deep}
A.~Alkhateeb, S.~Alex, P.~Varkey, Y.~Li, Q.~Qu, and D.~Tujkovic, ``Deep
  learning coordinated beamforming for highly-mobile millimeter wave systems,''
  \emph{{IEEE} Access}, vol.~6, pp. 37\,328--37\,348, 2018.

\end{thebibliography}
